\documentclass[12pt]{article}
\usepackage[centertags]{amsmath}
\usepackage{amsfonts}
\usepackage{amssymb}
\usepackage{amsthm}
\newcommand{\beq}{\begin{equation}}
\newcommand{\eeq}[1]{\label{#1}\end{equation}}
\newcommand{\bea}{\begin{eqnarray}}
\newcommand{\eea}[1]{\label{#1}\end{eqnarray}}

\newcommand{\Tr}{{\rm Tr}\,}

\begin{document}
\setlength{\topmargin}{-1cm} \setlength{\oddsidemargin}{0cm}
\setlength{\evensidemargin}{0cm}
\begin{titlepage}
\begin{center}
{\Large \bf The Three Dimensional Dual of 4D Chirality}

\vspace{20pt}

{\large M. Porrati$^a$ and L. Girardello$^b$}

\vspace{12pt}

$^a$ Center for Cosmology and Particle Physics\\
Department of Physics\\ New York University\\
4 Washington Place\\ New York, NY 10003, USA

\vspace{12pt}

$^b$ Dipartimento di Fisica, Universit\`a di Milano-Bicocca\\ and \\ INFN,
Sezione di Milano-Bicocca\\ Piazza della Scienza 3, Milano I-20126, Italy

\end{center}
\vspace{20pt}

\begin{abstract}
Chiral gauge theories can be defined in four-dimensional Anti de
Sitter space, but AdS boundary conditions explicitly break the chiral
symmetry in a specific, well defined manner, which in turns
results in an anomalous Ward identity. When the 4D theory admits a
dual description in terms of a 3D CFT, the 3D dual of the broken
chiral symmetry is a certain double-trace deformation of the CFT,
which produces the same anomalous chiral Ward identities that obtains in the 4D bulk theory.
\end{abstract}

\end{titlepage}

\newpage
\section{Introduction}
The recent renewed interest in $AdS_4/CFT_3$ duality has a twofold origin. On the one hand, ref.~\cite{abjm} (see also~\cite{bl}) constructed an explicit 3D superconformal field theory with Chern-Simons terms, which is holographically dual to M theory compactified on $AdS_4 \times S^7/Z_k$; on the other hand, string or M theory compactifications to $AdS_4$ are a crucial step in the search for a landscape~\cite{s} of realistic vacua with positive cosmological constant~\cite{kklt}. Knowing the holographic dictionary between string theory vacua on $AdS_4$ with matter content compatible with the Standard Model and their dual $CFT_3$ would be, therefore, of great practical as well as theoretical value. Any such string vacua must contain chiral families. String compactifications to $AdS_4$ spaces that
contain Fermions charged under a chiral gauge group are indeed known, to various degrees of explicitness~\cite{adhl,af,dvz}.

Even before any detailed quest for realistic gauge groups and family structure, the very presence of a chiral gauge
group in the ``bulk'' $AdS_4$ description raises a simple question, that also arises in studying the supersymmetric Standard Model on
$AdS_4$~\cite{gkrrs}:
\begin{center}
{\em What is the 3D, CFT dual of 4D Chirality?}
\end{center}
Aim of this note is to answer that question. We will confine ourselves to a toy model containing one or several 4D
chiral Fermions, charged under a $U(1)$ field. This simple setting is sufficient to answer our question and also to discuss the effect of gauge anomalies in $AdS_4$.

 We start with an analysis of chirality in $AdS_4$, in Section 2. There, we present a self-contained derivation of the known result that $AdS_4$ boundary conditions explicitly break chirality and  we also find the crucial Ward identity that will allow us to discover the CFT dual of chiral rotations. This Ward identity encodes the invariance of the on-shell 4D action under certain changes of the boundary values of the $U(1)$ vector and the chiral Fermions. Its detailed form depends on specific properties of the 4D bulk action such as locality, a canonical kinetic term etc.

The 3D meaning of the Ward identity
is described in Section 3, where we show that a 4D chiral rotation manifests itself in the dual 3D CFT as a particular double-trace deformation~\cite{w01} of the theory. Section 4 concludes the paper with a short and sketchy discussion of the holographic correspondence beyond the leading order of the large-$N$ expansion. Beyond leading order, bulk gauge anomalies may appear that make flat-space theories inconsistent. In Section 4, we argue that the explicit breaking of chirality due to $AdS_4$ boundary conditions allows for a Green-Schwarz anomaly cancelation mechanism, which blurs somewhat the distinction between anomalous and anomaly free theories in Anti de Sitter space. We thus expect that the 3D dual of an anomalous theory is qualitatively similar to the dual of an anomaly free one.
\section{Chirality and Its Breaking in $AdS_4$}
We begin this Section with a more careful discussion of what chirality is in an $AdS_4$ background.

As pointed out in~\cite{al} and recently emphasized in~\cite{gkrrs}, chirality is broken by the conformal boundary of $AdS_4$. The physical reason is that when Weyl Fermions reflect off the $AdS_4$ boundary, their helicity flips; a more formal reason goes as follows.
The action of a chiral Fermion coupled to gravity is
\beq
S={i\over 2} \int d^4 x e \psi^*_{\dot\alpha}\bar{\sigma}^{\dot\alpha \alpha\; a}\stackrel{\leftrightarrow}{D}_m \psi_\alpha e^m_a + ....
\eeq{1}
Here we used the convention of Wess and Bagger~\cite{wb} for spinor indices, we denoted with $D_m$ the spin covariant derivative, with $e^a_m$ the vierbein, and with $e$ its determinant. So, we raise and lower both dotted and undotted spinor indices as $\psi^\alpha=\epsilon^{\alpha\beta}\psi_\beta$, $\psi_\alpha=\epsilon_{\alpha\beta}\psi^\beta$, $\epsilon_{12}=-\epsilon^{12}$. In particular, $\bar{\sigma}^{\dot\alpha \alpha\; a}=\epsilon^{\dot\alpha\dot\beta}\epsilon^{\alpha\beta}\sigma_{\beta\dot\beta}^a$.

By varying the action and setting all fields on shell, we are left with a boundary term. Near the conformal boundary, we
call $z$ the radial direction and $x^\mu$, $\mu=0,1,3$ the boundary coordinates; then, the $AdS_4$ metric can be written as $e^m_a=z\delta^m_a$. Notice that we chose to identify the Lorentz index of the radial coordinate as $a=2$. With this choice a bulk Weyl spinor decomposes in a very simple way under the boundary Lorentz group. Namely, its real and imaginary parts do not mix: each one transforms as a real doublet of $SL(2,R)$.

Noticing that $e=\det e_m^a = z^{-4}$, the boundary term becomes
\beq
\delta S = {i\over 2}\int d^3 x z^{-3}\left( \psi^*_{\dot\alpha}\bar{\sigma}^{\dot\alpha \alpha\; 2}\delta \psi_\alpha
-\delta\psi^*_{\dot\alpha}\bar{\sigma}^{\dot\alpha \alpha\; 2} \psi_\alpha \right).
\eeq{2}

Since in 3D there is no difference between dotted and undotted indices, we will drop the distinction henceforth. This identification eliminates the distinction between $\sigma$ and $\bar\sigma$. We also rescale the (massless) Weyl Fermion as $\psi_\alpha=z^{3/2} \phi_\alpha$ to get
\beq
\delta S = {i\over 2}\int d^3 x \left( \phi^*_{\alpha}\sigma^{\alpha \beta\; 2}\delta \phi_\beta
-\delta\phi^*_{\alpha}\sigma^{\alpha \beta\; 2} \phi_\beta \right).
\eeq{3}
Since the Fermionic equations of motion are of first order, not all components of the complex field $\phi_\alpha$ are independent. Half of them are
independent external sources and half are determined by the equations of motion and by specifying the behavior of $\phi_\alpha$  inside $AdS_4$~\footnote{In Euclidean signature, this means regularity at all points inside $EAdS_4$; in Lorentzian signature, this means an appropriate behavior at past infinity (in global coordinates) or on the past horizon (in the Poincar\'e patch).}.
The source transforms covariantly under the 3D Lorentz group $SL(2,R)$. In our spinor basis this means
\beq
\phi_\alpha = {1\over \sqrt{2}}e^{-i\theta/2}(\chi_\alpha + i\eta_\alpha),
\eeq{4}
where the source $\chi$ is kept fixed while the ``conjugate variable'' $\eta$ is free. So, when varying the action we must set to zero only the change
$\delta S / \delta \eta$; we thus find the equation
\beq
\delta S = {\delta S \over \delta \phi_\beta}{\delta \phi_\beta \over \delta \eta_\alpha}\delta \eta_\alpha +
{\delta S \over \delta \phi^*_\beta}{\delta \phi^*_\beta \over \delta \eta_\alpha}\delta \eta_\alpha =0,
\eeq{5}
which gives
\beq
ie^{-i\theta/2}{\delta S \over \delta \phi_\beta} - i e^{i\theta/2}{\delta S \over \delta \phi^*_\beta}=0.
\eeq{6}
This equation must be identically zero since otherwise it would imply a local linear constraint between the source $\chi$ and the conjugate variable $\eta$. This would be inconsistent with the true equation for $\eta$, which follows from writing $\delta S / \delta \chi$ in terms of $\chi$ and $\eta$.
The variation~(\ref{3}) does not make eq.~(\ref{6}) vanish identically. To achieve this, we must add a boundary term
\beq
S_B=\int d^3x (C \phi_\alpha \phi_\beta - C^* \phi^*_\alpha \phi^*_\beta )\epsilon^{\alpha\beta}.
\eeq{7}
By demanding that eq.~(\ref{6}) vanishes identically we find $C={1\over 4}\exp(i\theta)$.

Crucially, with our choice of spinor basis, 4D chirality acts by rotating $\phi \rightarrow \exp(i\varphi)\phi$; therefore, the boundary term~(\ref{7}) explicitly breaks chirality. To regain an invariance of the action
$S+S_B$, one must also shift the external parameter $\theta$ as $\theta \rightarrow \theta - 2\varphi$.  Definition~(\ref{4}) then implies that $\chi$ and $\eta$ are left invariant by the combined action of the chiral rotation and the shift in $\theta$.

We are studying a theory where the chiral $U(1)$ is a gauge symmetry with an associated Abelian gauge field $A_m$, which transforms as
$A_m \rightarrow A_m + \partial_m \varphi$. Since the gauge action is scale invariant in 4D, no rescaling is needed to define the boundary value of the gauge field, which we call $A_\mu$. The action is invariant under a local chiral transformation, when accompanied by a shift in $\theta$ and a gauge transformation in $A_\mu$ ($S_T=S+S_B$). The on-shell action depends only on the 3D boundary values of the dynamical fields; therefore, we
obtain a relation which can be interpreted as a Ward identity in a 3D theory:
\beq
S_T(e^{i\varphi}\phi,e^{-i\varphi}\phi^*,A_\mu + \partial_\mu \varphi, \theta -2\varphi)=S_T(\phi,\phi^*,A_\mu,\theta).
\eeq{8}
It is suggestive to think of $\theta$ as a ``boundary axion;'' this intuition will be made more precise later.
In infinitesimal form, after using the bulk equations of motion and performing an integration by part in $dz$, eq.~(\ref{8}) becomes a boundary identity:
\bea
\int d^3x \delta\varphi(x)\left[i{\delta S_T \over \delta \phi(x)}\phi(x) - i{\delta S_T \over \delta \phi^*(x)}\phi^*(x) -\partial_\mu {\delta S_T \over \delta A_\mu(x)} \right]
 =&& \nonumber \\ \int d^3x \delta\varphi(x)\left[{i\over 2}e^{i\theta}\phi_\alpha(x) \phi_\beta(x) + {i\over 2}e^{-i\theta} \phi^*_\alpha(x) \phi^*_\beta(x) \right]\epsilon^{\alpha\beta}.&&
\eea{9}
The explicit form of $\delta S_T / \delta \phi(x)$ found using Eqs.~(\ref{3},\ref{7}) is
\beq
{\delta S_T \over \delta \phi(x)}=-{1\over 2} \phi^*(x) + {1\over 2} e^{i\theta}\phi={i\over \sqrt{2}}e^{i\theta/2}\eta ,
\eeq{10}
thus eq.~(\ref{9}) becomes equivalent to
\beq
\partial_\mu {\delta S_T \over \delta A_\mu(x)} = -i \phi^*_\alpha \phi_\beta \epsilon^{\alpha\beta}.
\eeq{11}
This equation is given in terms of only boundary values of fields and holds for the action computed on shell. If the 4D theory admits a holgraphic
dual, eq.~(\ref{11}) becomes a Ward identity for the dual 3D CFT. Even though in form it may seem identical to the Ward identity of vector-like
gauge transformations, its meaning is different here because of a different identification between sources and fields.

The simplest example of a vector theory is obtained by using two spinor fields,
$\phi^\pm_\alpha$, of opposite $U(1)$ charge $\pm q$. In this case one can add to the canonical bulk action of two chiral Fermions a charge-preserving boundary term
\beq
S_{VB}={1\over 2}\int d^3 x (C\phi^+_\alpha \phi^-_\beta - C^* \phi^{-*}_\alpha \phi^{+*}_\beta )\epsilon^{\alpha\beta}.
\eeq{v1}
By choosing $|C|=1$ one ensures that the bulk action $S_V$ plus the boundary term, $S_{VT}=S_V+S_{VB}$, only depends on the linear combinations
\beq
\psi\equiv{1\over \sqrt{2}} (\phi^+ - C^*\phi^{-*}), \qquad \psi^*\equiv{1\over \sqrt{2}} (\phi^{+*} - C\phi^{-}),
\eeq{v2}
which thus play the role of canonical coordinates. The conjugate variables are then
\beq
{\delta S_{VT} \over \delta \psi(x)}= {1\over \sqrt{2}}(\phi^{+*}+ C \phi^{-}),
\qquad {\delta S_{VT} \over \delta \psi^*(x)}=-{1\over \sqrt{2}}(\phi^{+}+ C^* \phi^{-*}).
\eeq{v3}
Now, instead of eq.~(\ref{11}) one gets
\beq
\partial_\mu {\delta S_{VT}\over \delta A_\mu(x)} = -i q(\phi^{+*}_\alpha \phi^+_\beta -\phi^{-*}_\alpha \phi^-_\beta)\epsilon^{\alpha\beta}.
\eeq{v4}
This is a standard non-anomalous Ward identity
\beq
 \partial_\mu {\delta S_{VT}\over \delta A_\mu(x)}= -iq{\delta S_{VT} \over \delta \psi(x)} \psi (x) +iq {\delta S_{VT} \over \delta \psi^*(x)}\psi^*(x).
\eeq{v5}

In the chiral case instead, the link between source $\chi$ and field $\eta$ is given by eq.~(\ref{4}). Thus, the Ward identity~(\ref{11}) is
\beq
\partial_\mu {\delta S_T \over \delta A_\mu(x)} = -{i\over 2} \chi_\alpha \chi_\beta \epsilon^{\alpha \beta}
-{i\over 2} \eta_\alpha \eta_\beta \epsilon^{\alpha \beta}.
\eeq{12}
It can be rederived also using eq.~(\ref{4}). By definition the fields $\chi$, $\eta$ in eq.~(\ref{4}) are invariant under the symmetry $\phi \rightarrow \exp(i\varphi) \phi$, $\theta \rightarrow \theta - 2\varphi$, $A_m \rightarrow A_m + \partial_m \varphi$. Under a generic change in $\theta$ and  $A_m$ keeping $\chi,\eta$ fixed, the 4D action transforms as
\beq
\delta S = \int d^4x e \left[ F^{mn} \partial_m \delta A_n+ {1\over 2}\psi^*_{\dot\alpha}\bar{\sigma}^{\dot\alpha \alpha\; a}\psi_\alpha e^m_a (2\delta A_m+\partial_m \delta\theta)\right].
\eeq{12a}
Using the bulk equations of motion and integrating by part we find the change of the on-shell action
\beq
\delta S = \int d^3x \left[ F^{2\mu}\delta A_\mu +  {1\over 2}\psi^*_\alpha\sigma^{\alpha \beta\; 2}\psi_\beta \delta\theta\right].
\eeq{12b}
By specializing this equation to the case of the chiral $U(1)$ symmetry $\delta A_\mu = \partial_\mu \delta \varphi$, $\delta \theta = -2 \delta \varphi$, which leaves the action invariant, we recover again Ward identity~(\ref{12}).

This derivation makes clear that the boundary term~(\ref{7}) is not unique. We can add to it any boundary term which depends on $\chi$ only. This non-uniqueness reflects in the dual 3D CFT in the possibility to add certain double-trace deformations without spoiling conformal invariance.

Identity~(\ref{12}) follows from (and conversely implies) an anomalous (operatorial) conservation equation for the chiral current $J^\mu$:
\beq
\partial_\mu J^\mu = -{i\over 2} \chi_\alpha \chi_\beta \epsilon^{\alpha \beta} -{i\over 2} \eta_\alpha \eta_\beta \epsilon^{\alpha \beta}.
\eeq{13}
Though similar in form to eq.~(\ref{12}), in eq.~(\ref{13}) $J^\mu$ and $\eta$ are operators, while $\chi$ is still the external source for $\eta$. Since $\eta$ has dimension $3/2$ --up to corrections to be discussed later-- the last term in the right-hand-side of eq.~(\ref{13}) is an internal (i.e. operator-valued) anomaly, while the first is an external anomaly, present only when sources are non-vanishing.

In this Section, we started from a 4D chiral field theory on $AdS_4$ and found a Ward identity that encodes the effect of chiral rotations. An effect survives even though chirality is explicitly broken by the boundary conditions. In the next Section, we will find a deformation of the 3D dual CFT, which yields again eq.~(\ref{12}). Quite explicitly, we will thus find the 3D dual of a 4D chiral rotation.
\section{The $CFT_3$ Story}
A reasonable guess for the deformation is that it is a double trace one~\cite{w01}. As customary in the case of CFTs admitting $AdS$ duals, we will imagine that the theory admits some sort of large $N$ expansion. To be concrete we shall assign adjoint indices to the fundamental fields of the CFT.
In this Section, we will work mostly at leading order in the $1/N$ expansion. At this order, it is most convenient to describe double trace deformations
in terms of the effective action $\Gamma(\eta,A_\mu,\theta)$~\cite{mue}, whose Legendre transform in $\eta$ gives the free energy of the CFT, which here we called  $S_T(\chi,A_\mu,\theta)$.

At large $N$~\footnote{Throughout this section, $S_T$ and $\Gamma$ are appropriately rescaled to remain finite in the $N\rightarrow \infty$ limit while $\chi$ is
rescaled to have a VEV $O(N^0)$. For an adjoint theory, this means that when the kinetic term of the fundamental
fields $\Phi$ is $\sim N\Tr \partial_\mu \Phi \partial_\mu \Phi$, the free energy is $N^2 S_T$ and
$\chi \sim N^{-1} \Tr P(\Phi)$, where $P$ is a polynomial with $N$-independent coefficients.}, the most general marginal double trace deformation in $\eta$ is
\beq
S_T(\chi)= -\Gamma(\psi) + i \int d^3x \left(B\chi_\alpha \psi^\alpha +{A\over 2} \chi_\alpha \chi_\beta \epsilon^{\alpha\beta} + {C\over 2} \psi^\alpha \psi^\beta \epsilon_{\alpha\beta} \right),
\eeq{14}
where $S_T$ is evaluated at the stationary point in $\psi$
\beq
{\delta \Gamma \over \delta \psi^\alpha}=  iB \chi_\alpha +iC\psi^\beta\epsilon_{\beta\alpha}.
\eeq{15}
The deformation must be marginal to preserve conformal invariance. Compared with~\cite{w01,mue}, here we have an extra term, quadratic in the source $\chi$, which changes the relation between the variable $\psi$ and the field $\eta$. It is of course  still true that $\eta \sim \delta S_T / \delta \chi$; precisely
\beq
{\delta S_T \over \delta \chi_\alpha} = i\eta^\alpha = -iA\epsilon^{\alpha\beta}\chi_\beta - i B \psi^\alpha.
\eeq{16}

Notice that because of definition~(\ref{4}), $\chi$ is invariant under chiral rotations; thus, eq.~(\ref{8}) can be re-written as
\beq
S_T(\chi,A_\mu + \partial_\mu \varphi, \theta -2\varphi)=S_T(\chi,A_\mu,\theta).
\eeq{17}
or, in infinitesimal form
\beq
-\partial_\mu {\delta S_T \over \delta A_\mu(x)} =2{\delta S_T\over \delta \theta(x)}.
\eeq{18}
This identity, together with eq.~(\ref{16}) allows us to rewrite eq.~(\ref{12}) as
\beq
2{\delta S_T\over \delta \theta(x)}={i\over 2}(1+A^2) \chi_\alpha \chi_\beta \epsilon^{\alpha \beta} -{i\over 2} B^2 \psi^\alpha \psi^\beta \epsilon_{\alpha\beta}  + iAB \chi_\alpha \psi^\alpha .
\eeq{19}
The explicit form of eqs.~(\ref{12},\ref{19})'s right hand side is determined by the whole 4D action. That is the point where holography provides us with additional properties beyond those valid in any 3D CFT. So, eqs.~(\ref{12},\ref{19}) are holographic while eq.~(\ref{18}) is of course generic.

In definition~(\ref{14}), all dependence on $\theta$ comes through the deformation parameters $A,B,C$. In order to reproduce the Ward identity~(\ref{12}), these parameters can be chosen to be local and to depend on $\theta(x)$ only, not on its derivatives. This property and eq.~(\ref{14}) give the identity
\beq
{\delta S_T\over \delta \theta(x)}= {i\over 2} \partial_\theta A\chi_\alpha \chi_\beta \epsilon^{\alpha \beta} +{i\over 2} \partial_\theta C \psi^\alpha \psi^\beta \epsilon_{\alpha\beta}  + i\partial_\theta B \chi_\alpha \psi^\alpha .
\eeq{20}
Substitution into eq.~(\ref{19}) finally gives a set of ODEs that determine the $\theta$ dependence of $A,B,C$.
\beq
2\partial_\theta A = 1 +A^2, \qquad 2\partial_\theta B = AB, \qquad 2\partial_\theta C = -B^2 .
\eeq{21}
The solution obeying the obvious initial condition $A=C=0$, $B=\pm 1$ at $\theta=0$ is
\beq
A=\tan(\theta/2),\qquad B=\pm{1\over \cos(\theta/2)},\qquad C= -\tan(\theta/2).
\eeq{22}
The coefficients $A,B,C$ are defined in the interval $0\leq \theta < \pi$. At $\theta=\pi$ they exhibit a singularity, which signals
that the role of source and field in eq.~(\ref{4}) is interchanged. This phenomenon is similar to the behavior of other
CFTs under double-trace deformations~\cite{kw}.

An easy concrete example of the general structure outlined above is that of a quadratic effective action. To define it properly we continue the Lorentzian action to Euclidean signature. The Euclidean Fermions $\psi$ and $\chi$ are now complex and the effective action is
\beq
\Gamma(\psi)= {1\over 2}\int {d^3k  \over (2\pi)^3}\psi^\alpha (-k) \left(a |k|^{-1}{/\!\!\!k}_{\alpha\beta} + ib\epsilon_{\alpha\beta}\right)\psi^\beta(k),
\eeq{23}
where $a\neq 0$, $b$ are arbitrary real constant.
Its form is fixed by demanding a correct Lorentz-invariant continuation to Minkowski signature, by conformal invariance and  by the conformal weight of $\psi$, $\Delta=3/2$. The constant $a$ can be rescaled to $a=1$ while $b$ is a marginal deformation that preserves conformality~\cite{mue,w01}. The freedom to change $b$ parallels the freedom to add certain
boundary terms to the 4D action we mentioned at the end of Section 2.

The stationary point condition~(\ref{15}) gives
\beq
\psi^\alpha = i {B\over a^2 +(C-b)^2} \left[a |k|^{-1} {/\!\!\!k}^{\alpha\beta} - i(C-b)\epsilon^{\alpha\beta}\right]\chi_\beta(k),
\eeq{23a}
while eq.~(\ref{16}) gives the conjugate variable $\eta$ as
\beq
\eta_\alpha = i\left( X |k|^{-1} {/\!\!\!k}_\alpha^{\;\;\beta} -iY\delta_\alpha^\beta\right)\chi_\beta(k).
\eeq{23c}
The free energy $S_T$ is then trivial to compute using definition~(\ref{14}):
\beq
S_T={1\over 2}\int {d^3k  \over (2\pi)^3}\chi_\alpha (-k) \left(X|k|^{-1}{/\!\!\!k}^{\alpha\beta}+ iY \epsilon^{\alpha\beta}\right)\chi_\beta(k).
\eeq{24}
The coefficients $X$ and $Y$ depend on $\theta$; their explicit form becomes, after some elementary trigonometry
\beq
X=-{a\over F(\theta)}, \qquad Y= -{d\over d\theta} \log F(\theta), \qquad F(\theta)= {1\over 2}(a^2 + b^2 -1) \cos \theta + b \sin \theta +
{1\over 2} (a^2 + b^2 +1).
\eeq{24a}
A straightforward calculation shows that eq.~(\ref{20}) is satisfied for arbitrary $a\neq 0$, $b$.
\section{The $CFT_3$ Story at Order $1/N^2$}
Up to now we have studied chiral rotations and their 3D dual in the large $N$ limit, which corresponds to the tree level of the bulk theory in $AdS_4$. Yet most of our conclusions are valid beyond leading order in the $1/N$ expansion.
One such thing is eq.~(\ref{14}), which can be recast as an exact functional integral identity~\cite{ar,egpr}. By using the same notations as in the first footnote of Section 3, the CFT Lagrangian is $L=N^2 O(\Phi)$, where the single-trace operator $O(\Phi)$ is normalized to have finite expectation value in the large $N$ limit. The identity reads
\beq
e^{N^2 S_T(\chi)}= \int [d\psi]\int [d\Phi] \delta[ \psi - N^{-1}\Tr P(\Phi)]e^{-\int d^3x L(\Phi) + iN^2 [B\chi_\alpha \psi^\alpha +(A/2) \chi_\alpha \chi^\alpha + (C/2) \psi^\alpha \psi_\alpha]}.
\eeq{25}
By performing the constrained functional integral in $[d\Phi]$ one obtains $\exp(-\Gamma)$ and at large $N$ the functional integral in $[d\psi]$ reduces to a Legendre transform, hence this formula reduces to eq.~(\ref{14}) in the large $N$ limit.

Ward identity~(\ref{13}) too remains valid beyond the leading order. What changes is that the scaling dimensions of
$J^\mu$ and $\chi$ receive nonvanishing corrections at order $1/N^2$. That those corrections are non-vanishing follows from a calculation in the dual 4D theory~\cite{rr}. To wit, the boundary term~(\ref{7}) modifies the propagator of the bulk Fermion $\psi$. In turn, this modification changes the one-loop self energy of the bulk photon $A_m$, inducing a finite, nonzero mass term. If the theory contains $n$ Fermions $\psi_i,..,\psi_n$ with chiral $U(1)$ charges $q_1,..,q_n$, the square mass is~\cite{rr}
\beq
m^2 = \sum_{i=1}^n q_i^2 {4g^2\over 3(4\pi)^2}L^{-2}.
\eeq{26}
In AdS/CFT the 4D $U(1)$ coupling constant $g$ is $O(1/N)$ and the dimension of the current $J^\mu$ is $(\Delta +1)(\Delta -2)=m^2L^2$ so $J^\mu$ acquires a dimension $\Delta \approx 2 + O(N^{-2}\sum_{i=1}^n q_i^2)$.

Notice that the anomalous dimension $\Delta -2$ is nonzero also when $\sum_{i=1}^n q_i^3=0$ i.e. when the gauge theory is anomaly free. In fact, the explicit dependence of the free energy $S_T$ on $\theta$ somewhat blurs the difference between anomalous and non-anomalous gauge theories in $AdS_4$.

As we mentioned earlier, $\theta(x)$ shifts under gauge transformations and couples to the boundary term~(\ref{7}), hence the name ``boundary axion.'' When extended into the bulk and to $O(1/N^2)$, $\theta$ acquires a kinetic term and becomes a bona fide axion,  which can be used to cancel the $U(1)$ gauge anomaly by the 4D Green-Schwarz mechanism~\cite{gs,w84}.
Namely, the $U(1)$ gauge field mass~(\ref{26}) implies that the axion possesses an effective kinetic term, which below the energy scale $1/L$ assumes the standard form
\beq
\int d^4 x \sqrt{-g} {m^2\over 2 g^2} g^{mn}(A_m + \partial_m \theta/2)(A_n + \partial_n \theta/2).
\eeq{27}
The 4D gauge anomaly is canceled by adding to the bulk action the local term
\beq
{1 \over 32\pi^2} \sum_{i=1}^n q_i^3\int d^4 x \sqrt{-g} \theta F_{mn}\tilde{F}^{mn}.
\eeq{28}
After a rescaling  $\theta= 2g/m \theta_c$, which canonically normalizes the $\theta_c$ kinetic term, the dimension-five
operator $\theta_c F\tilde{F}$ in~(\ref{28}) is multiplied by the nonrenormalizable coupling constant $\Lambda^{-1}\equiv |\sum_{i=1}^n q_i^3| L /8\pi\sqrt{\sum_{i=1}^n q_i^2}$, which sets the cutoff of the theory to $\Lambda$. A consistent $AdS_4$ theory most probably must come from dimensional reduction of a ten dimensional string background (or an 11D M-theory background). The existence of Kaluza-Klein modes implies that the 4D theory changes at energies $O(1/L)$. When the anomaly coefficient $|\sum_{i=1}^n q_i^3|$ is smaller than  $\sum_{i=1}^n q_i^2$, the cutoff $\Lambda$ implied by the Green-Schwarz term is higher than $O(1/L)$. When $|\sum_{i=1}^n q_i^3|\ll\sum_{i=1}^n q_i^2$
the 4D theory can make sense by itself, even without the KK mode completion, up to an energy scale parametrically higher than $1/L$. So, at least in some cases, it appears that an anomalous theory in $AdS_4$ does not exhibit features qualitatively different from an anomaly free one. In particular, in either case, their 3D holographic dual possesses a current obeying eq.~(\ref{13}) and acquiring an anomalous dimension at $O(1/N^2)$.
\subsection*{Acknowledgments}
We would like to thank A. Zaffaroni for participating in the early phase of this work and for many interesting discussions. We thank R. Rattazzi and  M. Redi for valuable discussions and for sharing with us
their computation of gauge boson masses in $AdS_4$ prior to
publication; we thank D. Malyshev, F. Marchesano, H. Ooguri, L. Rastelli, G. Shiu, A. Uranga and F. Zwirner
for valuable discussions and correspondence. We thank the Galileo Galilei
Institute for Theoretical Physics, Florence for its
hospitality during completion of this work. L.G would
like to thank the CCPP and the NYU Physics
Department and M.P. would like to thank The Scuola Normale Superiore, Pisa for their hospitality at various stages during the  completion of this work. M.P. is supported in part by NSF grant PHY-0758032, and
by ERC Advanced Investigator Grant n.226455 {\em Supersymmetry,
Quantum Gravity and Gauge Fields (Superfields)}; L.G. is
supported in part by INFN, by MIUR contract
2007-5ATT78-002, and by the European Commission RTN program MRTN-CT-
2004-005104.

\end{document}